\documentclass[oneside,a4paper]{article}

\usepackage{hyperref}
\usepackage{amsmath}
\usepackage{amssymb}
\usepackage{graphicx}
\usepackage{subfig}
\usepackage{authblk}
\providecommand{\keywords}[1]{\textbf{\textit{Keywords:}} #1}

\title{Flexible skylines, regret minimization and skyline ranking: a comparison to know how to select the right approach}
\author{Vittorio Fabris}
\affil{Politecnico di Milano\\
Milan, Italy\\
\href{mailto:first.last@polimi.it}{vittorio.fabris@mail.polimi.it} }
\date{}

\begin{document}
\maketitle
\begin{abstract}
Recent studies pointed out some limitations about classic top-$k$ queries and skyline queries. Ranking queries impose the user to provide a specific scoring function, which can lead to the exclusion of interesting results because of the inaccurate estimation of the assigned weights. The skyline approach makes it difficult to always retrieve an accurate result, in particular when the user has to deal with a dataset whose tuples are defined by semantically different attributes. Therefore, to improve the quality of the final solutions, new techniques have been developed and proposed: here we will discuss about the flexible skyline, regret minimization and skyline ranking approaches. We present a comparison between the three different operators, recalling their way of behaving and defining a guideline for the readers so that it is easier for them to decide which one, among these three, is the best technique to apply to solve their problem.
\end{abstract}

\keywords{top-$k$ queries, skyline queries, flexible skyline, regret minimization, skyline ranking\\}

\section{Introduction}
Nowadays, databases contain lots of data. The user who looks for some information aims to retrieve the best solutions that the database contains with respect to some preferences. However, providing in the exact way the most suitable tuples for the case is not always a simple task. \\
Based on their request, the deployment of \textit{multi-criteria decision-making} tools helps users to retrieve the best available results in the database. For this reason, users should select the most appropriate approach according to the current situation, so that they are guaranteed to get the finest answer for their query. In fact, during the last decades several different techniques have been proposed in literature. Remarkable ones are the \textit{top-k query}, the \textit{lexicographical approach} and the \textit{skyline query}.\\\\
Top-$k$ queries extract the top-$k$ objects from the database based on a \textit{scoring function}. The score of an object is based on its characteristics, which means that the values of its attributes contribute to the overall score amount. The most simple and commonly used scoring function is the weighted-sum function: each attribute contributes in a different way according to some delineated weights that have been set a priori, usually by the users. The tuning weights operation needs to be very precise because even a small variation in the assigned values can deeply change the final ranking. Most of the times users do not precisely know what effort to assign to the attributes, compromising the final result and retrieving something that can be far from the \textit{real best} result that they could have extracted from the database. The same parameter tuning process could be repeated multiple times to average the retrieved rankings, but this means to waste time and resources. \\
What’s more, the weighted formula approach can also mix different non-commensurable attributes into the same instance: values that cannot be compared are included in the same computation and this leads to an incoherent result, going against the principle for which in data extraction processes the retrieved information should always be accurate and understandable to the users \cite{Fayyad_Piatetsky-Shapiro_Smyth_1996}.\\
For this reason, lots of studies proposed new different algorithms that deal with the inefficiency \cite{Re_Dalvi_Nilesh} and the imprecisions carried by this method, tackling them through different techniques, as explained also in \cite{surveytopk}. The performances of this kind of approach have improved, but the drawbacks pointed out before are still a considerable component that affects the problem.\\
In the lexicographical approach it is established a priori a linear priority order among the attributes. This means that between two tuples, the one with the best value in the highest priority attribute is preferred to the other. This happens even if all the other attributes of the first tuple show worse values, having no possibility to compensate. This highlights the fact that, as it happens when using the weighted-formula, the lexicographical approach deals with the so called missed opportunities \cite{cm20}: it can happen that the real best result is not provided to the users.\\
Skyline queries are based on the \textit{Pareto approach}, introduced in \cite{TheSkylineOp}, whose goal consists into selecting all the possible non-dominated tuples in the database. In particular, a tuple $t_1$ is dominated by a tuple $t_2$ if the values of all the attributes of $t_2$ are better or equal than the ones of $t_1$, but at least one is better. However, since this method does not make a distinction among the retrieved tuples in the solution, it is up to the users to select the one that best suits their purposes. When the cardinality of the database is huge and the attributes are difficult to compare to each other, it is very likely that the proposed solution is heavy from the point of view of the number of retrieved tuples. Of course, this makes hard for the users to select the best case to solve their problem, as directly documented also in \cite{SKYLINENEW}.\\\\
The limitations described before among all the different approaches lead the research community to improve the existent techniques, developing new algorithms and procedures. \\The main purpose of this survey is to focus on some of these techniques, explaining how they work and their inner properties. Then, a comparison between them is made so that the readers can decide which is the approach that best suits their situation. \\
The exploited techniques are:
\begin{itemize}
\item Flexible skyline;
\item Regret minimization;
\item Skyline ranking.\\
\end{itemize}
The rest of the paper is organized as follows. Section 2 presents the flexible skyline and the concept of $F$-dominance, with the related properties and algorithms that characterize them; Section 3 recalls the notions about the regret minimization approach, while Section 4 introduces to the world of the skyline ranking queries. Then, a comparison between the three approaches is made in Section 5, and the final considerations are made in Section 6 to conclude the paper.\\

\section{Flexible Skyline}
As it has been pointed out before, it is difficult for the users to guess and set the right value for the weights in a scoring function. What’s more, considering only ranking functions instead of skyline techniques or vice versa does not lead to precise results. For this reason a mix between the two approaches helps to reduce the drawbacks that one technique carries when it is used alone, compensating the final result and retrieving a better representation of what the users expects from their query. \\
This is the reason why the authors in \cite{CM1} introduced the notion of \textit{flexible skyline} queries: taking into account the difference of importance between the attributes, users do not have to set the precise values for their weights, but it is sufficient that they formulate some constraints to state and bound the relationships that exist between them. This allows us to introduce a higher flexibility in the system, which gives the final result more chances to contain the tuple that the users are looking for and, at the same time, to avoid the risk of providing the whole database as a solution. \\
For a more detailed introduction to $F$-skylines, the reader should refer to \cite{CM1}, while here are only reported its main concepts to allow a global understanding of their main characteristics. \\
The main node about this type of queries is the concept of $F$-dominance: a tuple $t_1$ $F$-dominates another tuple $t_2$ when $t_1$ is always better than or equal to $t_2$ according to all the scoring functions in a certain family $F$. The set $F$ contains multiple scoring functions, which are all taken into account so that even an incomplete or approximated first evaluation of the attribute's weights is covered by one of them. In this framework, two operators have been introduced in order to characterize the set of non-$F$-dominated tuples (i.e., $ND$) and to address the tuples which are potentially optimal according to some function in $F$ (i.e., $PO$). \\\\
A practical example can help the reader to better understand the concepts that have just been presented. We consider the situation where the players of a football team want to define who is the most valuable player among them. We consider their statistics of goals and assists during their last winning season. The classic skyline, as shown in Figure 1, demonstrates that Harry, Sam, Andrea and Samuel contributed significantly for the final win of the championship, while Thomas, Robert and Alex have been excluded from the final result because they have been dominated by their teammates. Then, we consider the same dataset and apply the Flexible Skyline approach: $F$ is the set of linear scoring functions of the form $MostValuablePlayer = w_{1} * Goals + w_{2} * Assists$ such that $w_{1}>w_{2}$. The $PO$ result that contains the set of potentially optimal tuples is composed by Harry and Sam (Thomas is still dominated). As we were expecting, the result size is smaller than the one of the classic skyline applied at the beginning, and we can better identify who should finally run for the \textit{Most Valuable Player} title in the team.
\begin{figure}[ht!]
\centering
\includegraphics[width=90mm]{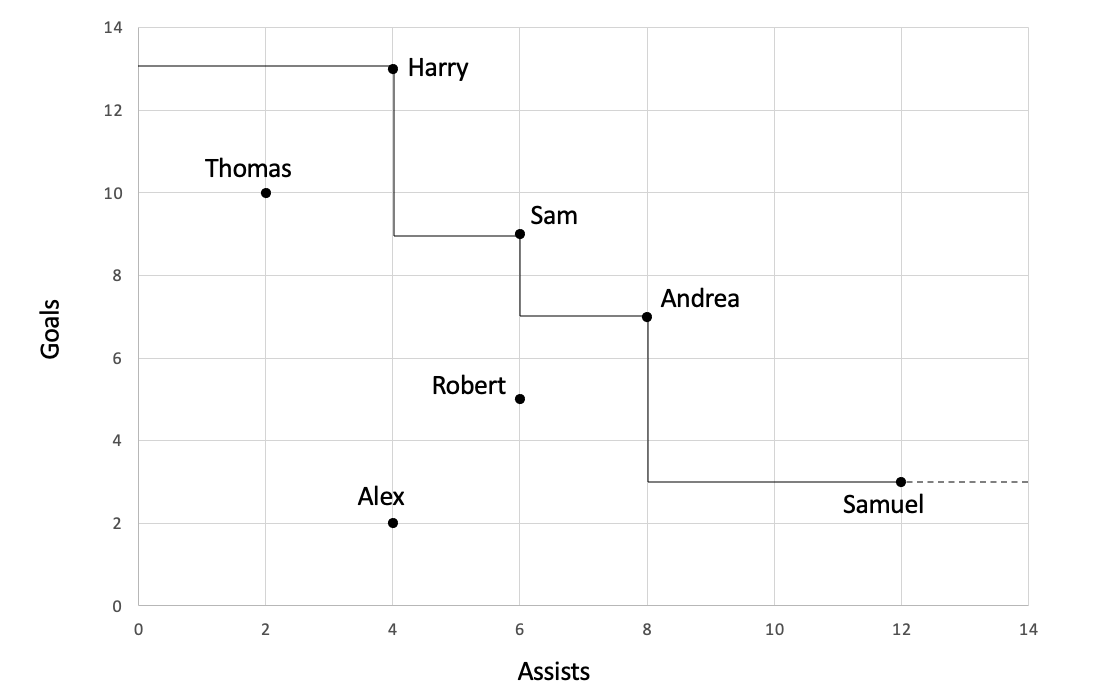}
\caption{Skyline of the most valuable players of a football team} 
\label{football}
\end{figure}

\noindent
$F$-skylines may seem to be equivalent to other developed approaches available in literature. In the following lines, we want to clarify their way of behaving, comparing them to other known techniques. Some examples where the $F$-dominance approach is applied are treated too, so that an idea about how they can be exploited can be useful for the user's experience.\\
As well as $F$-skylines, prioritized skylines \cite{pskyline} reduce the size of the result following the user's preferences. However, the way the attributes’ preference is expressed is different, because instead of a set of constraints that regulates the importance between each other, a strict priority between them exists. For this reason, the final tuple extraction shows a lower flexibility into the selecting process. Some of the tuples that could fulfill the user’s requirements are not selected because of the current order of the attributes, and no compensating operation is performed in order to recover them.\\
The same words can be spent for the \textit{restricted skyline} introduced by \cite{RESTRICTEDSKYLINE}, where, thanks to the application of the Weak Pareto Dominance condition and due to pruning actions and prime cuts \cite{RESTRICTEDSKYLINE2} performances, the tuples can be removed if they do not fit with the user preferences with respect to a partial order instantiated between the attributes. \\
We would like to highlight the fact that the constraints that are set for the flexible skyline are not made directly on the specific domain that an attribute can have, reducing the research field of the constrained dimension, as it happens in \cite{RESTRICTEDSKYLINE3} and in \cite{RESTRICTEDSKYLINE4}, where a condition to reduce the size of the operands involved in the computation is expressed to deploy an algebraic optimization operation that reduces the overall costs of the procedure. $F$-skylines share with the classic skylines the capability to provide an overall view of interesting results, without limiting the domain of a dimension, but allowing the data analyst to focus on specific parts of the skyline, depending on the user preferences. The framework is able to express preferences like “attribute $a_1$ is more important than attribute $a_2$, but no more than twice as important” \cite{cm_main}. Therefore, this approach is also different to the ones which assume to give the attributes a weight based onto a probabilistic function, such as in \cite{CM42}. It is now easy to see that neither the \textit{$\varepsilon$-skyline} \cite{FLEXIBLE1} or the \textit{thick skyline} \cite{FLEXIBLE3} deal with the problem in the same way that the $F$-skyline does. While the former two control the cardinality of the final result, which can be regulated with respect to the user needs, they do not care in the same way about the user’s attributes' preference as the latter one.
Dynamic skylines \cite{papadias_tao,FLEXIBLE6} exploit a different perspective, because while they both deal with multiple scoring functions, they consider them for different purposes: dynamic skylines deal with build-up tuples based on the original data and transformed through a certain function; flexible skylines deal with an infinite set of scoring functions and the concept of $F$-dominance includes them all at the same time when points are evaluated.\\\\
Concerning $F$-skylines, the authors in \cite{CM1,ciacciamart2,cm_main} introduced different algorithms implementing the $ND$ and $PO$ operators with the aim of extracting the flexible skyline with respect to the set of scoring functions $F$. The reader is invited to check Resource \cite{cm_main} to read the details about the time and space complexities of the exploited procedures. \\
In \cite{ciacciamart3} the authors develop another algorithm based on the concept of $F$-Dominance, with the aim of sourcing the top-$k$ tuples from a database. In this scenario, not all the weights for the attributes are known and therefore more than a single scoring function needs to be considered in order to be sure that users get what they are looking for. FSA, which is based on the Fagin’s Algorithm (FA) \cite{CM3_6_fagin} and on the Threshold Algorithm (TA) \cite{CM_7_faginTA}, proves to be able to keep the cardinality of the result to a reasonable value and outperforms other similar reference procedures, gaining a performance level that can be compared to the one of the classical top-$k$ queries. \\
Mouratidis and Tang introduced the concept of \textit{uncertain top-k queries (UTK)} \cite{mouratidis_tang1}, where they decide to face the problem of the estimation of the weights by extending the usual specific vector to a broader region. Two algorithms are proposed and they are both based on the notion of $F$-dominance. In this scenario, where weights are not well-defined because we are talking about their area of pertinence and not about their specific value, it can happen that a record might dominate another due to its wider boundaries (not limited to a fixed value, as it happens in the traditional way) and then we’re allowed to prune the useless tuples. In \cite{mouratidistang2} it is highlighted the fact that the problem belongs to the \textit{RP} group, since its preference input is a region $R$ and gives as output the top-$k$ results for the query (as points $P$).\\\\
In conclusion, we can say that the user that wants to retrieve information in a flexible way, interacting with the problem by setting (even partially) the weights for the parameters involved in it, retrieving a final outcome which considers all the dimensions of the problem and that is presented in a form that reminds the one of the classic skyline, can find the concept of $F$-dominance very interesting for his application.\\

\section{Regret Minimization}
\begin{figure}[h!]
\centering
\includegraphics[width=90mm]{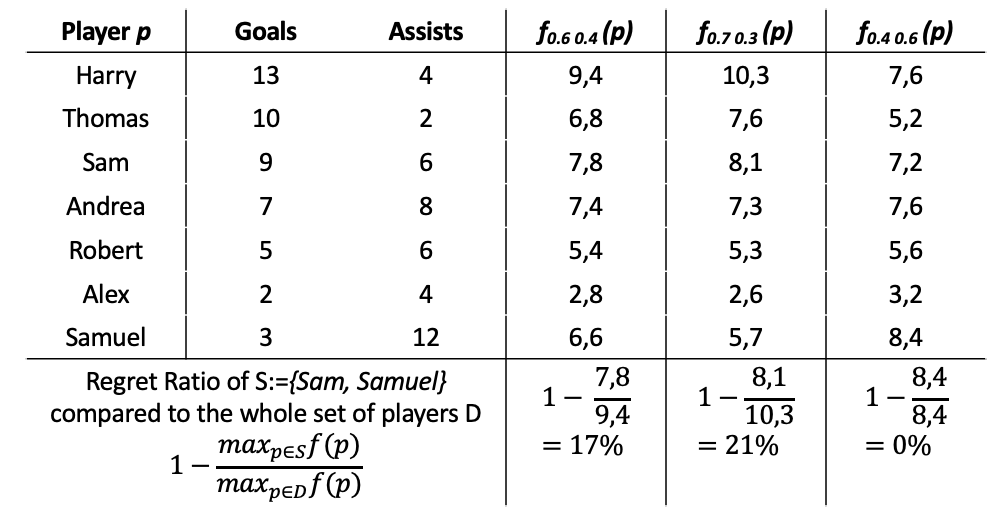}
\caption{Example of the regret ratio results for a subset of the football players problem. In this scenario, the last function guarantees that with the selected subspace the user will not be regretful at all because the solution contains the best tuple possible in the database.} 
\label{regretmin}
\end{figure}

The regret minimization approach has been introduced by Nanongkai et al. \cite{nanongkaietal} to reduce the drawbacks given from both classic top-$k$ and skylines techniques highlighted in previous sections, while trying to take by its side the benefits that the same methods can offer, such as the possibility to provide a controlled number of interesting tuples without stressing the user to define the perfect scoring function. \\
The regret minimization performs a computation to extract a representative subset of the database so that for any preference vector that users would set when deploying other techniques, the top ranked points in the subset are a good approximation of the top ranked items in the whole database, fulfilling the users' expectations and interests. \\
Therefore, it is possible to deal with something which is smaller than an eventual result proposed by the application of a classic skyline algorithm. In particular, given a database $D$ and the desired output size $k$, the final goal is to find a set of $k$ tuples that minimizes the \textit{maximum regret ratio} \cite{nanongkaietal}. This means that, while having no information about the users' preferences, the algorithm tries to satisfy the most part of them reducing at its minimum the percentage of the unsatisfied ones. A user is not regretful with a given subset if his/her \textit{regret ratio} is close to zero, since the highest utility in the subset is close to the best utility in the dataset. \cite{chesterthomo} introduces the concept of \textit{k-regret-minimization}, relaxing the initial condition where the regret ratio denoted how far from the best score in the dataset is the best score in the subset: the preference scores of the best tuple in the subset is then compared with the users top-k choice.\\\\
The \textit{regret minimization query (RMS)} is scale-invariant, because the regret ratio does not change if we scale each point of the dataset for a certain factor. In fact, the computation of the regret is performed as a ratio, and therefore the scaling operation has no effect on the final output. What’s more, the RMS is said to be \textit{stable}. When a \textit{junk point} (i.e., a point that shows never into user’s preferences and will be never considered into the RMS generated subset because it does not have an high utility for any considered function) is added or removed from the dataset, the regret amount does not change.\\
Research in this field tried to maximize the possible achievable rewards thinking about guaranteeing the best possible regret ratio in order to satisfy the users’ preferences. However, if we want to go a little bit more into the details, in \cite{REGRET17,efficient_kregret_query} the authors pointed out that if the database has $p$ points, it is not possible to guarantee a maximum regret ratio better than $\Omega(p^{-\frac{2}{d-1}})$, where $d$ is the dimension of the problem.\\
Searching for an optimal solution for RMS, which is to find the minimum size set for which a certain regret ratio $\varepsilon$ is guaranteed, \cite{chesterthomo} demonstrated that it was a NP-hard problem, and then \cite{REGRETNEW4} and \cite{REGRETNEW} supported this conclusion for dimensions $d > 2$.\\\\
We do not spend so much time to group here the algorithms that are used to solve the RMS problem, as it is not the main purpose of this paper. The reader can find a good reference in \cite{REGRETNEW3}, where different state-of-the-art methods are presented and reviewed. They describe and compare different procedures, covering a full range from the ones that interactively deal with the user (e.g., \cite{REGRET17,REGRETNEW5,REGRETNEW6}) to the ones that greedily approximate solutions for the high-dimensional scenarios, where deterministic solutions are not available (e.g., \cite{REGRETNEW3,REGRETNEW2}), and comment on their way of behaving and their performances with respect to classic skylines or top-$k$ queries.\\\\
In the end, just to focus a little bit more on the pragmatic aspects of this technique, we can reason about some possible applications of it. We can recall the same scenario mentioned in the previous Section \ref{football}. Through Figure \ref{regretmin}, we can show to the reader some practical results of the problem, tackled with the regret minimization approach, highlighting the regret ratios obtained considering some specific functions.\\ 

\noindent
Otherwise, we can think about other possible applications that well-fit with this kind of approach, e.g., someone who is doing a research or a survey on a particular subject and who wants to see which are the preferences of a certain population in a specific field. Researchers may want to choose this method because it aims at finding the best result so that every user preference is covered. This is guaranteed by the fact that the extracted results are not based on a specific scoring function, but they try to satisfy all the people involved in a broader way. The results retrieved from this analysis can be used in a second moment, to define marketing specifications so that a product can be designed to be sold to a larger amount of people and guaranteeing a higher final income.\\

\section{Skyline Ranking}
\begin{figure}[ht!]
	\subfloat[Skyline of subspace $\left \{ Goals, Dribblings \right \}$]{
		\begin{minipage}[1\width]{
		0.40\textwidth}
		\centering
		\includegraphics[width=0.99\textwidth]{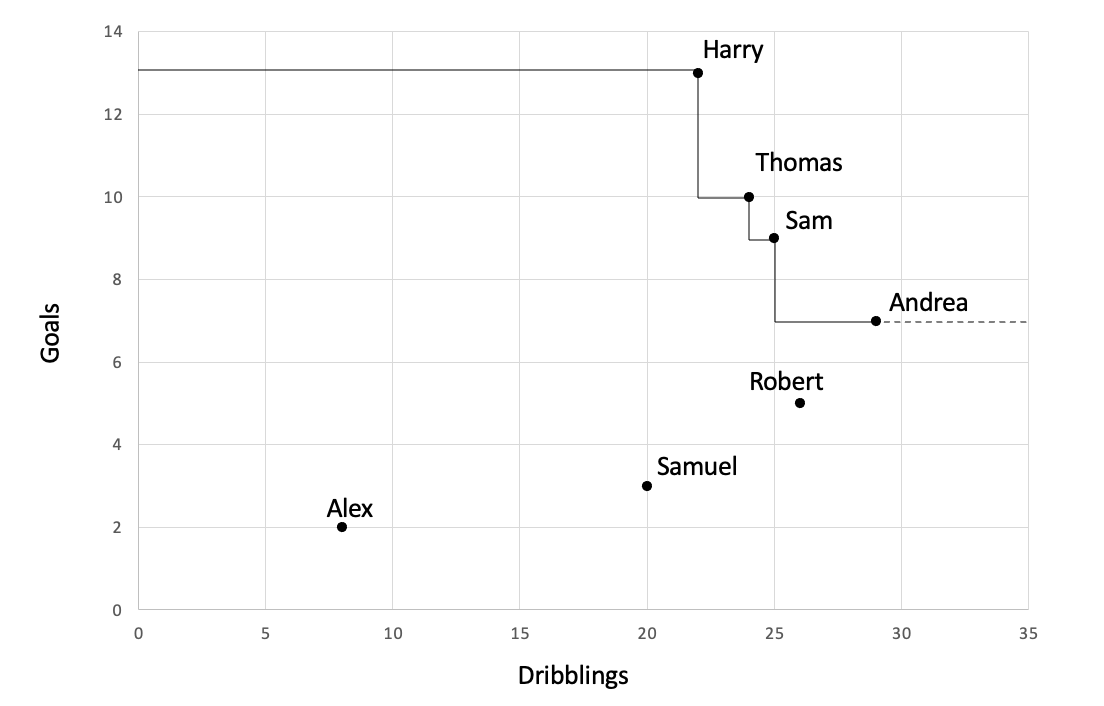}
		\end{minipage}}
	\hfill
	\subfloat[Skyline graph for the $\left \{ Goals, Dribblings \right \}$ subspace ]{
		\begin{minipage}[1\width]{
		0.3\textwidth}
		\centering
		\includegraphics[width=0.99\textwidth]{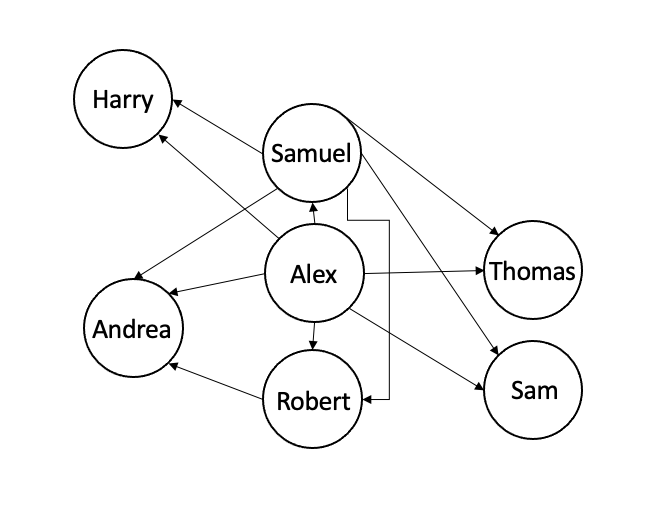}
		\end{minipage}}
	\hfill
	\subfloat[Final ranking]{
		\begin{minipage}[1\width]{
		0.27\textwidth}
		\centering
		\includegraphics[width=0.99\textwidth]{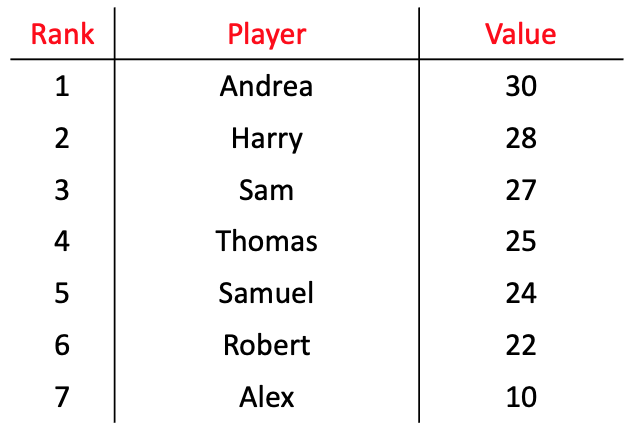}
		\end{minipage}}
	\caption{}
	\label{footballdrib}
\end{figure}

The capability of proposing to the user an ordered list of the most satisfactory results for his research has always been a fundamental task to solve and whose performances should be improved as much as possible. Ranking data in a good way allows the user to spot and use in a quicker way the needed information. \\
We can think about how ranking techniques have successfully been applied in the web search field \cite{Google_pagerank} or simply when we’re doing a keyword research in a database \cite{SKYLINEORD7}. \\
As the name suggests, skyline ranking aims to deploy an hybrid approach between the skyline and the top-$k$ ordering techniques. In fact, the goal of skyline ranking is to extract the interesting skyline points from a dataset and rank them based on the utilities they provide. The more points in different subspaces a point dominates, the more important it is and therefore is very likely to find it in the final ranking. Subsets are intended as \textit{subsets of attributes} of the global space. So, this meaning is basically different to the one of subsets of points contained in the dataset, as for example it has been exploited in G-Skyline queries \cite{GSkyline} or their variants (e.g., FGSky queries \cite{FLEXIBLE2}).\\
\cite{VLACHOUvazirgiannis} considers that the interestingness of a point is higher if it dominates many other important skyline points in different subspaces, and it is even more relevant when the dominated ones also dominate other points in other subspaces. This concept is very close to the one of the skyline frequency \cite{frequencyORDER}, where the more times a point shows in a subset’s skyline, the more probability it has to be in the main skyline that defines the set. The \textit{SKYRANK} method presented in \cite{VLACHOUvazirgiannis} considers this idea and improves its technical behavior, as it does not care so much about the so called extreme points (i.e., the skyline points of the one-dimensional subspaces), producing a very nice result. The skyline graph is a useful tool that allows us to order all of the interleaving relationships between subspaces and skyline points, so that even in the absence of a preference function given by the user this approach perfectly works. \\\\
We can say that the classic $SKYRANK$ approach is basically different with respect to other techniques that hybridize the skyline and top-$k$ queries approach. In fact, approaches like \cite{mouratidistang3}, where we can find the attempt of controlling the cardinality of the output and give to the users the most relevant results for their research based on a certain defined preference vector, do not completely fit with the $SKYRANK$’s main idea.
However, the basic version of it can be extended to handle top-$k$ preference skyline queries when the user preferences are available. \\
\textit{Skyline cubes} \cite{SKYLINEORD30} help in doing the job, allowing the algorithm to efficiently select only the possible non-empty subsets of the global dataset saving time in the computation and enhancing then the overall performance. \\
\cite{lujensenzhang} takes advantage of partitioning the global dataset and introduces the notion of \textit{size constrained skyline queries}, which are able to retrieve $k$ desirable points from the $d$-dimensional dataset: the \textit{skyline ordering} approach creates a skyline-based partitioning of the points and instantiates an order among such partitions. It has been proved that skyline ordering enhances the general performances of the constrained skyline query \cite{SKYLINEORDNEW}. Then, when selecting the $k$ points of the result, this approach progressively considers the points of the partitions in the order that is in force between them. When the $k$ points are retrieved, the last partitions are pruned. Points are not evaluated one by one according to a specific mapping function and they are neither taken all together selecting the best ones in order to maximize an overall score that shows the greatness of the retrieved result, because it would be computationally expensive. So, points are organized in batches and then ranked one after the other, balancing the two approaches just presented.\\
As it has been mentioned above, visiting all the points of the skyline introduces some drawbacks from the performance point of view. For the same reason, the \textit{IR-style} ranking mechanism presented in \cite{SKYLINEORDNEW4} improves the initial naïve behavior, which implied the full scan of the set, by discarding as soon as possible the points that for sure will not be included in the final result to present to the user. In addition, \cite{SKYLINEORDNEW4} shows that even if without making a proper division of the dataset into multiple subspaces, the ranking idea where the points which are dominated the less by the others have an higher value is preserved.\\\\
To understand the steps that this kind of approach requires to be working, we can set up a basic example. Considering the situation presented in Figure \ref{football}, we complete the scenario introducing other data that allow us to extend the number of subspaces of the problem. In addition to \textit{Goals} and \textit{Assists} we introduce the \textit{Dribblings} performed by each player. Figure \ref{footballdrib} shows the relations between the different players, and provides a different skyline with respect to the one analyzed in the previous example. The computation of the skyline graph allows to determine the dependences between dominated and dominating points: points that have only outgoing edges are dominated by the ones they are pointing to, while points that have only incoming edges are dominating points and are belonging to the skyline of that subspace. After the skyline graph construction, link-based ranking techniques are applied in order to assign the score to each vertex (i.e., skyline point). For example, as it happens with web pages, we can think that the edge of the skyline graph transfers some importance to the pointed items: skyline points that dominate many other skyline points, which in turn dominate other points in some subspaces, at the end are highly ranked. Therefore, we can explain why the value amount of Samuel is lower than the one of Sam and Harry: while the former belongs only to one of the two skylines of the analyzed subspaces, the latter two are not dominated in both of the two skylines. This justifies the different values of their final scoring and their final positions in the ranking. \\

\noindent
In conclusion, skyline ordering techniques and algorithms like $SKYRANK$ can be very helpful when we want to mitigate top-$k$ and skylines' drawbacks, but we still want to be sure to include all the dimensions of the problem into our investigation and to benefit of the pros of the two mentioned approaches. The final ranking that we obtain from the technique's deployment improves our capabilities of analyzing the scenario where we are operating. This allows us to have a clear and ordered representation of the best ways to solve our problem and select the best item we are looking for.\\

\section{Comparison between the different techniques}
Now that we overviewed the three different techniques, recalled their way of behaving and main properties, we can define in which situations and in which modalities the user can choose to use one method instead of the other. In fact, while they have been shown to overclass more classic operators such as top-$k$ and skyline, some discriminants can be pointed out to decide that one is preferable to the other.\\
Table \ref{table1} helps to schematically sum-up and highlight the differences, the pros and the cons of each technique that will be discussed more in the details in this section. Then, users will be able to address their research towards a certain direction instead of another.

\begin{table}[ht]
\begin{center}
\begin{tabular}{||c||c|c|c||}
\hline
\hline
& $F$-skyline & Regret Minimization & Skyline Ranking \\
\hline\hline
Flexible input & \checkmark & \checkmark & \checkmark \\
\hline
Mix top-$k$ \& skyline approaches & \checkmark & \checkmark & \checkmark \\  
\hline
Weight-based & \checkmark &  &  \\
\hline
Multiple scoring functions & \checkmark &  &  \\
\hline
User interaction & \checkmark & possible & possible \\
\hline
Scale invariant &  & \checkmark & \checkmark \\
\hline
Stable &  \checkmark & \checkmark & \checkmark \\
\hline 
Attribute order compensation & \checkmark & \checkmark & \checkmark \\
\hline
$k$ selection &  & \checkmark & \checkmark \\
\hline
Rank results &  &  & \checkmark \\ 
\hline
All-$d$ inclusive & \checkmark & \checkmark & \checkmark \\
\hline
Uncontrollable output size &  &  &  \\
\hline
Partial imprecision of output &  & \checkmark &  \\
\hline\hline
\end{tabular}
\end{center}
\caption{Main characteristics and properties of the studied operators. To have a full and complete explanation of the table please refer to Section 5, where parameters are depicted with respect to each one of the three analyzed techniques. The checkmark in a cell means that the property mentioned on the same row of the left-most column of the table is satisfied by the selected approach.}
\label{table1}
\end{table}

\noindent
As we have considered so far, all the three approaches improve the performances guaranteed by the top-$k$ and skyline operators. Features belonging to the two latter methods are considered at the same time, so that the drawbacks that characterize them when they are used singularly are mitigated and the advantages they carry are exploited at the same time. In addition, all the three surveyed techniques give the possibility to provide a flexible input that doesn't stress users to define into the details the precise values of the efforts to assign to the discriminants of their problem. \\
In particular, from this last point of view some considerations about the $F$-skyline operator can be made. When a comparison with ranking queries is made, while it increases the flexibility, being able to control the size of the result, they do not prevail on the other approach from the performance efficiency point of view. Even if it lacks a satisfying general method of comparison, by taking into account that the scoring function of the ranking queries approach is a single one while the results of the flexible skyline are based on a whole set, it has been shown that the top-$k$ query retrieves more easily the interesting tuples than the $F$-dominance-based approach: $F$-skylines pay the increased capability of returning interesting results at the price of an higher computational overhead with respect to ranking queries \cite{cm_main}.  \\\\
Moving to another issue, we can consider the situation where we want to express our preferences about properties (i.e., the attributes), because in our research we would like to concentrate on some aspects rather than some others. Therefore, we might look for a solution that allows us to personalize the effort that some components have in the final result, as it can be done when setting the values for the weights in the scoring function in the top-$k$ queries. As we have seen so far, flexible skylines can perfectly fit the situation too, because we can set constraints on weights according to our preferences. For this reason, as it happens for the two other approaches (i.e., regret minimization and skyline ranking), there is not a proper scoring function, but we can say that a weight-based method is applied to guarantee to have not a lack of personalization in the solution. In fact, the concept of $F$-dominance is applied with respect to a whole group of scoring functions, not only a single one. On the other hand, when the users have no idea about what are the ratios between the attributes, they might prefer to go for a method that provides them a solid solution simply by extracting from the dataset the points that have an higher score compared to the others. Therefore, regret minimization and skyline ranking are the indicated techniques to apply.\\
However, if the users are firmly decided to apply the regret minimization or the skyline ranking approaches for a particular reason, we mentioned the fact that it is possible to deploy some specific algorithms that would allow them to be included into the tuples' extraction process in an interactive way.\\\\
A brief analysis of some inner properties can be useful, so that the users can know whether or not they are allowed to perform some kind of operations. For example, while the regret minimization and skyline ordering operators are scale invariant, the $F$-skyline does not behave in the same manner. In fact, as there have been set constraints on weights between attributes, changing the scale (and then the attributes' score) can compromise the validity of the relations that were decided at the beginning. On the other side, all the three techniques are valid from the point of view of the stability. In fact, the same thought is valid for all of them: when new junk points are added to the dataset, they will not be included in the final result, keeping the size of the solution on the same level.\\
All the discussed approaches demonstrate an attitude to compensate the order among the attributes when they have to propose the final result. This means that they overcame the problem of the top-$k$ queries where the fixed preference imposed a situation where valid tuples were left out because of the strict relation imposed a priori. This behavior remarks another time the high flexibility of these methods, that are careful to include all the possible interesting solution to the user.\\\\
As it happens in classic top-$k$ queries, users may want to visualize a precise number of tuples in the retrieved solution. Therefore, regret minimization and skyline ordering are the two approaches they should consider. $F$-skyline doesn't include $k$ selection as its inner feature, it may be included in a second moment deploying specific algorithms to select the exact number of points. However, we would like to stress the fact that this is not a property which is included in the basic version of the flexible skyline approach. In addition, if we want to get the tuples sorted and ordered so that we can retrieve a ranking with the best points of the dataset, then we should go for the skyline ordering operator.\\
The three techniques allow us to have an overall view of interesting results, which is not always the case, as it happens in classic top-$k$ queries where some areas of the $d$-dimensional space are not considered enough because the weight of the attributes that characterizes them is not enough high to be relevant for the final result.\\ \\
What's more, the skyline classic uncontrollable output's size result is greatly mitigated, because the $k$ set specifies exactly the number of proposed solutions it wants to be retrieved from the algorithms, while the constraints imposed by the flexible skyline and the effect of the $F$-dominance relationship draw a narrower area to include a smaller amount of tuples. \\\\
In conclusion, users can think about the precision that they can get from the tuples' extraction process. $F$-skyline and skyline ranking guarantee to include the best tuples available in the dataset, while the regret minimization approach may only approximate the real best result that they can get form the operation. For this reason, users should evaluate the effort necessary to perform an exact computation on the available points that the former two methods can assure them and the estimated regret they have to deal with when applying the third technique, reducing the computational costs.\\

\section{Conclusions}
In this paper we have briefly reviewed the way of behaving and the main properties of three approaches: $F$-skyline, regret minimization and skyline ordering. Their aim is to overcome the limitations given by the classic methods such as top-$k$ queries and skyline queries, proposing a different manner to deal with the problem enhancing the global achievable performance of the process. We have then compared the different techniques, trying to define different scenarios where users can identify themselves so that they can choose which approach among the three is the best one to solve their problem. For example, the if the users know a priori the relationship that exists between the attributes of their problem then they should select the flexible skyline approach to set proper attribute boundaries, applying the concept of $F$-dominance and retrieving a more interesting skyline with respect to the classic one. Otherwise, if users do not want to influence the tuples extraction process defining a scoring function, they can decide to exploit the regret minimization approach. Finally, the possibility to obtain a ranking of the extracted tuples can be obtained when applying the skyline ordering technique.\\
Therefore, users are here provided of the necessary tools to decide how to solve their problem, without the risk of incurring in the limitations and drawbacks of classic skylines and top-$k$ queries. \\
What's more, once it has been decided to what approach go for, for a deeper study on which specific algorithm to deploy, the current paper also provides interesting references to articles and surveys about all the presented approaches.\\


\bibliographystyle{plain}
\bibliography{refs.bib}
\end{document}